\documentclass[twocolumn,prx,superscriptaddress,aps]{revtex4}
\usepackage{amsmath}
\usepackage{graphicx}
\usepackage{siunitx}
\usepackage{ulem}
\DeclareSIUnit\gauss{G}
\DeclareSIUnit\bohr{a_{B}}

\usepackage{hyperref}

\usepackage{color}
\definecolor{mygreen}{rgb}{0,0.5,0} 
\definecolor{mygrey}{rgb}{0.5,0.5,0.5} 
\definecolor{myred}{rgb}{0.75,0,0} 
\definecolor{myblue}{rgb}{0,0,0.75} 
\definecolor{mymagenta}{cmyk}{0,1,0,0.12} 
\definecolor{mycyan}{cmyk}{1,0,0,0.12} 
\definecolor{myorange}{rgb}{1,0.5,0}  
\definecolor{myviolet}{rgb}{0.5,0.3,1} 
\definecolor{mybrown}{rgb}{0.75,0.5,.5} 

\usepackage{ulem}

\newcommand{\DOne}{D$_1$}
\newcommand{\DTwo}{D$_2$}

\begin{document}
\newcommand{\microtrap}{microtrap~}
\newcommand{\microtraps}{microtraps~}

\newcommand{\mytitle}{A versatile and stable laser system for a single neutral atom microtrap}
\renewcommand{\mytitle}{{A versatile \microtrap system for neutral-atom quantum optics}}
\renewcommand{\mytitle}{{Manipulating and measuring single atoms in the Maltese cross geometry }}
\title{\mytitle}

\newcommand{\ICFO}{ICFO - Institut de Ci\`encies Fot\`oniques, The Barcelona Institute of Science and Technology, 08860 Castelldefels (Barcelona), Spain}
\newcommand{\ICREA}{ICREA - Instituci\'{o} Catalana de Recerca i Estudis Avan{\c{c}}ats, 08010 Barcelona, Spain}
\newcommand{\ZH}{State Key Laboratory of Modern Optical Instrumentation, College of Optical Science and Engineering, Zhejiang University, Hangzhou 310027,China}
\newcommand{\LENS}{LENS, European Laboratory for Non-linear Spectroscopy, Via N. Carrara 1, I-50019 Sesto Fiorentino, Firenze, Italy}

\author{Lorena C. Bianchet}
\affiliation{\ICFO}
\author{Natalia Alves}
\affiliation{\ICFO}
\author{Laura Zarraoa}
\affiliation{\ICFO}
\author{Natalia Bruno}
\affiliation{\LENS}
\author{Morgan W. Mitchell}
\affiliation{\ICFO}
\affiliation{\ICREA}

%
\begin{abstract}

We describe optical methods for trapping, cooling, and observing single $^{87}$Rb atoms in a four-lens ``Maltese cross'' geometry (MCG).  The use of four high numerical-aperture lenses in the cardinal directions enables efficient collection of light from non-collinear directions, but also restricts the optical access for cooling and optical pumping tasks.  We demonstrate three-dimensional atom localization with sub-wavelength precision, and present measurements of the trap lifetime, temperature and transverse trap frequency in this geometry. We observe a trap performance comparable to what has been reported for single-atom traps with one- or two-lens optical systems, and conclude that the additional coupling directions provided by the MCG come at little cost to other trap characteristics. 
\end{abstract}

\maketitle


Optical \microtraps at the focus of high numerical aperture (high-NA) imaging systems enable efficient collection, trapping, detection and manipulation of individual neutral atoms. These capabilities are exploited in several active topics in quantum optics and quantum technology, including strong single-atom effects on traveling-wave beams \cite{TeyNJP2009, ChinNC2017, AljunidPRL2009, TeyNP2008, LeongNC2016, SlodickaPRL2010}, higher-order interference of atoms \cite{KaufmanS2014, LesterPRL2018, KaufmanN2015} and Rydberg-atom-based quantum information processing \cite{SaffmanRMP2010} and quantum simulation \cite{BernienN2017, LabuhnN2016}. High-NA trapping systems may also enable strong modifications to radiation physics associated with sub-radiant states \cite{AsenjoGarciaPRX2017, PerczelPRL2017, GlicensteinPRL2020, RuiN2020}.   

\begin{figure}[h!]
\begin{center}
\includegraphics[width=0.6\columnwidth]{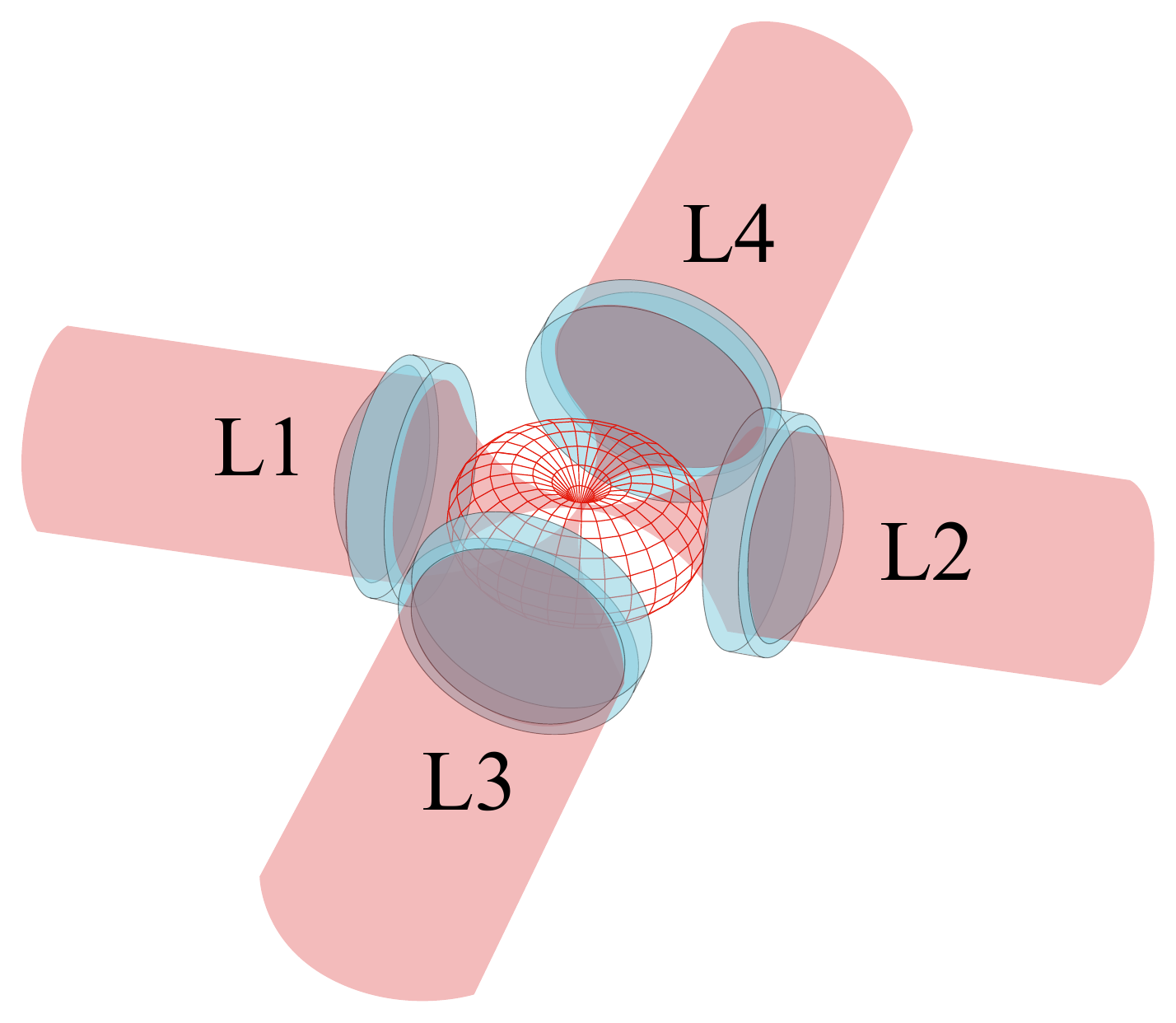} 
\caption{Illustration of the Maltese cross geometry, in which four high-NA lenses L1 to L4 are arranged along the cardinal axes. Red meshed structure at center shows the angular distribution of radiation from a vertically-polarized atomic transition.  The emission is strongest around the equator and thus efficiently collected by the four lenses. }
\label{fig:singletrap}
\end{center}
\end{figure}

The earliest experiments with neutral-atom \microtraps employed large vacuum systems and custom-designed optics \cite{SchlosserN2001}. More recent works have employed high-NA aspheric lenses in smaller vacuum systems, which has enabled experiments with high-NA optical access from two \cite{NogrettePRX2014, ChinNC2017} and four \cite{MartinezDorantesPRA2018, MartinezDorantesThesisBonn2016, BrunoOE2019, GlicensteinARX2021} directions. The latter scenario is known as the Maltese cross geometry (MCG) when the lenses are placed on the cardinal directions, as illustrated in \autoref{fig:singletrap}.  

The MCG both offers new opportunities and creates some new challenges in the design and operation of the trapping system.  In addition to boosting the total solid angle coupled to the atom \cite{BrunoOE2019}, the MCG promises to enable measurement of coherent, large-momentum-transfer scattering processes in disordered ensembles \cite{JenneweinPRL2016} and in atomic arrays \cite{AsenjoGarciaPRX2017}, for which strong sub-radiant effects are predicted.  The right-angle geometry is also predicted to enhance and modify the observable quantum correlations in resonance fluorescence \cite{GoncalvesARX2020}.  At the same time, the presence of four lenses necessarily reduces significantly the optical access to the trapping region, especially in the plane of the lenses.  This complicates some standard techniques such as a magneto-optical trap (MOT) operation in the usual six-beam configuration.  

 The article is organized as follows:  In \autoref{sec:Overview} we describe the experimental system, including MOT, far-off-resonance trap (FORT), and atomic fluorescence collection. In \autoref{sec:RightAngle} we study the ability of the four-lens system to produce a 1D optical lattice and make localized measurements of atomic occupation.  In \autoref{sec:Characterization} we characterize the trap lifetime, temperature, and trap frequencies.


\begin{figure*}[t]
\begin{center}
\includegraphics[trim= 0 100 0 0, width=0.98 \textwidth]{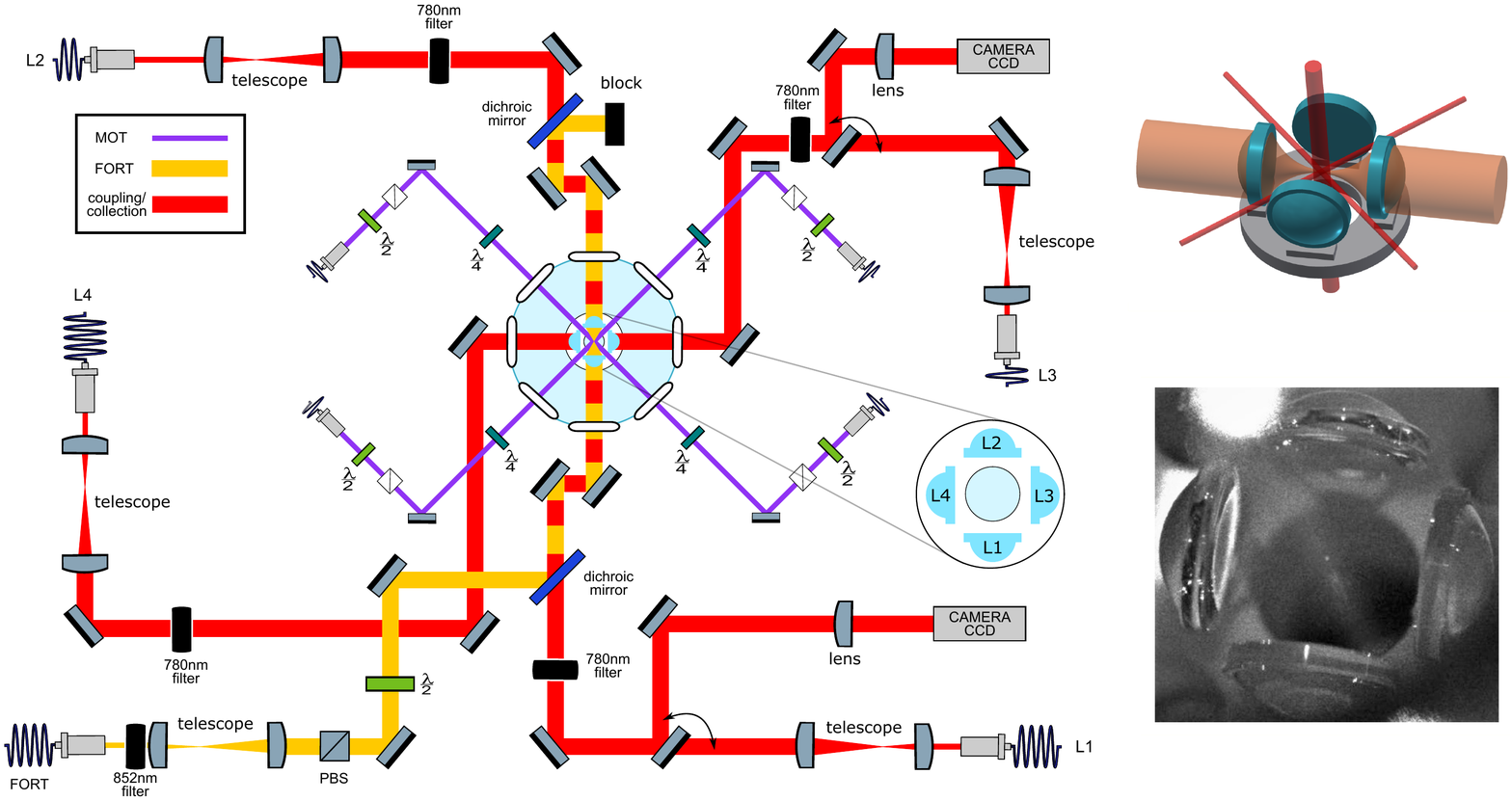}
\caption{Main elements of the optical setup. Left: Schematic of optical systems. Nearly all elements lie in a horizontal plane intersecting the trap center. In light blue: Four aspheric lenses (lens numbers are indicated in the inset) are located symmetrically around the geometric center of a ``spherical octagon'' vacuum enclosure with eight anti-reflection-coated windows. In purple: MOT beams, which pass through the gaps between the lenses; vertically-directed MOT beams passing through the trap center are not shown. In yellow: FORT beam, which is focused by L1 and re-collimated by L2. In red: Beams focused by the in-vacuum lenses for coupling light to and from the single atom. Red/yellow dashing: Coupling beams leading to L1 and L2 are combined on dichroic mirrors with the FORT beam, resulting in coaxial propagation. All four lenses can be used for fluorescence collection, which is sent to either a fiber or a CCD camera. Diagram symbols: PBS: polarizing beamsplitter, $\lambda/2$ ($\lambda/4$): half (quarter) waveplate,  filter: bandpass filters centered in the specified wavelength. Upper right: three-dimensional drawing of the four high-NA aspheric lenses, MOT beams (red) and FORT beams (peach). MOT beams have small diameters due to the limited space in between the lenses, leading to a cloud of cold atoms with a diameter of $\approx \SI{50}{\micro \meter}$. Lower right: photograph of the lens assembly in vacuum with MOT (bright spot) at center. See text for additional details.}  
\label{fig:schemeSetup}
\end{center}
\end{figure*}


\newcommand{\miniMOT}{mini-MOT}

%

\section{System Description} \label{sec:Overview}

The system employs a small MOT to collect and cool a cloud of $^{87}$Rb atoms from background vapor in an ultra-high vacuum enclosure, and load them into a FORT located within the MOT volume. The MOT and FORT centers are co-located at the center of a system of four high-NA lenses (NA=0.5) along the cardinal axes. A detailed description of the high-NA optics, assembly and characterization is given in \cite{BrunoOE2019}. Here we describe other critical elements of the trapping and cooling system, which is illustrated in \autoref{fig:schemeSetup}.

\subsection{MOT}\label{sec:MOT}

A small MOT is formed by six counter-propagating beams along three orthogonal axes in the standard configuration. Repumper light is on resonance with the $5S_{1/2}, F=1 \rightarrow 5P_{3/2}, F'=2$ transition. Cooler light is red-detuned from the $5S_{1/2}, F=2 \rightarrow 5P_{3/2}, F'=3$ transition by 6$\Gamma_0$, where $\Gamma_0 = 2 \pi  \times \SI{6.06}{\mega\hertz} $ is the {\DTwo} natural linewidth.  To pass cleanly between the \SI{1.2}{\milli \meter} gaps separating the lenses, the horizontally-directed beams are of \SI{0.7}{\milli\meter} diameter, whereas the vertical beams are of \SI{2.0}{\milli\meter} diameter.  Horizontal and vertical cooler beams have powers of \SI{20}{\micro\watt}  and  \SI{162}{\micro\watt}, respectively.  Repump light of \SI{150}{\micro\watt} is sent only in the downward vertical direction, to minimize scattered light. 
For the single-atom experiments described below, a MOT gradient of \SI{3.8}{\gauss\per\centi\meter} is used, to reduce the number of MOT atoms and resulting background fluorescence. 

\subsection{FORT} \label{sec:FORT}

The FORT is produced by a linearly-polarized \SI{852}{\nano \meter} beam with a power of  \SI{7}{\milli \watt} and a beam waist of \SI{1.85}{\milli \meter} at the aspheric lens position. 
The laser used to produce this beam is a distributed feedback (DFB) laser (Toptica Eagleyard EYPDFB0852) stabilized to the  $6S_{1/2}, F=4 \rightarrow 6P_{3/2}, F'=5$ Cs {\DTwo} transition by modulation transfer spectroscopy (MTS) \cite{EscobarAlvarezCoopEtAl2015}. 
The wavelength-scale size of the waist at focus creates a dipole micro-trap of few-\SI{}{\micro\meter\cubed} volume. In the presence of cooler light, e.g. if the MOT is on, light-assisted collisions (LACs) \cite{SchlosserN2001} rapidly remove any pairs of atoms in this small volume. In practice, this ensures the presence of no more than one atom in the trap.
The \SI{852}{\nano \meter} FORT wavelength is sufficiently far from resonance as to produce little scattering by the trapped atom, yet close enough that a single aspheric lens can be diffraction limited when focusing both it and the spectroscopic wavelengths \SI{780}{\nano \meter} ({\DTwo}) and \SI{795}{\nano \meter} ({\DOne}).  \SI{852}{\nano \meter} also coincides with the Cs {\DTwo} line, which is convenient for frequency stabilization and atomic filtering. To position the dipole trap midway between the two lenses, a shearing interferometer (SI) is used to measure the beam divergence before the input lens, and after the output lens, and to set the divergences to be equal and opposite. The same SI is used in this symmetric condition to check for aberrations.  For more details see \cite{BrunoOE2019}.

%

Within the Gaussian beam approximation, the FORT potential is
\begin{equation}
\label{eq:FORTPotential}
U_{\rm FORT}(r,z) = \beta P_{\rm FORT} \frac{2}{\pi w^2(z)} \exp[{-\frac{2r^2}{w^2(z)}}]
\end{equation}
where $r = \sqrt{x^2 + y^2}$ is the transverse radial coordinate, $z$ is the axial coordinate, $\beta \approx \SI{-6.39e-12}{\joule \meter\squared\per\watt}$ is the ground state light shift coefficient \cite{SimonCoopLightshift2017}, 
$P_{\rm FORT}$ is the power of the FORT beam, $w(z) \equiv w_{\rm FORT} \sqrt{1+ z^2/z_R^2}$ where $w_{\rm FORT}$ is the FORT beam waist,  and $z_R \equiv \pi w_{\rm FORT}^2/\lambda_{\rm FORT}$ is the Rayleigh length. 

In most circumstances, the atom's thermal energy is far less than the trap depth $k_B T_{\rm{atom}}\ll U_0 \equiv |U_{\rm{FORT}}(0,0)|$, where $k_{B}$ is the Boltzmann constant, and it is thus appropriate to use the harmonic approximation $U_{\rm{FORT}}(r,z) \simeq U_0 [-1+ 2(r/w_{\rm{FORT}})^2+(z/z_{R})^2]$. Based on the reported parameters of Volz \textit{et al.} \cite{VolzLP2007}, where a single atom trap with similar characteristics and FORT light wavelength is described, we predicted a value of $w_{\rm FORT} = \SI{1.6}{\micro\meter}$. For this value of  $w_{\rm FORT}$, the transverse and axial trap frequencies are then  $\omega_{\rm{r}}=\sqrt{4U_{0}/m_{\rm 87}w_{\rm{FORT}}^2} \approx\SI{56}{\kilo\hertz}$ and $\omega_{\rm{z}}=\sqrt{2U_{0}/m_{\rm 87}z_{R}^2} \approx\SI{6.7}{\kilo\hertz}$, respectively, where $m_{\rm 87}$ is the $^{\rm{87}}$Rb mass.

\begin{figure}[t]
\begin{center}
\includegraphics[width=1\columnwidth]{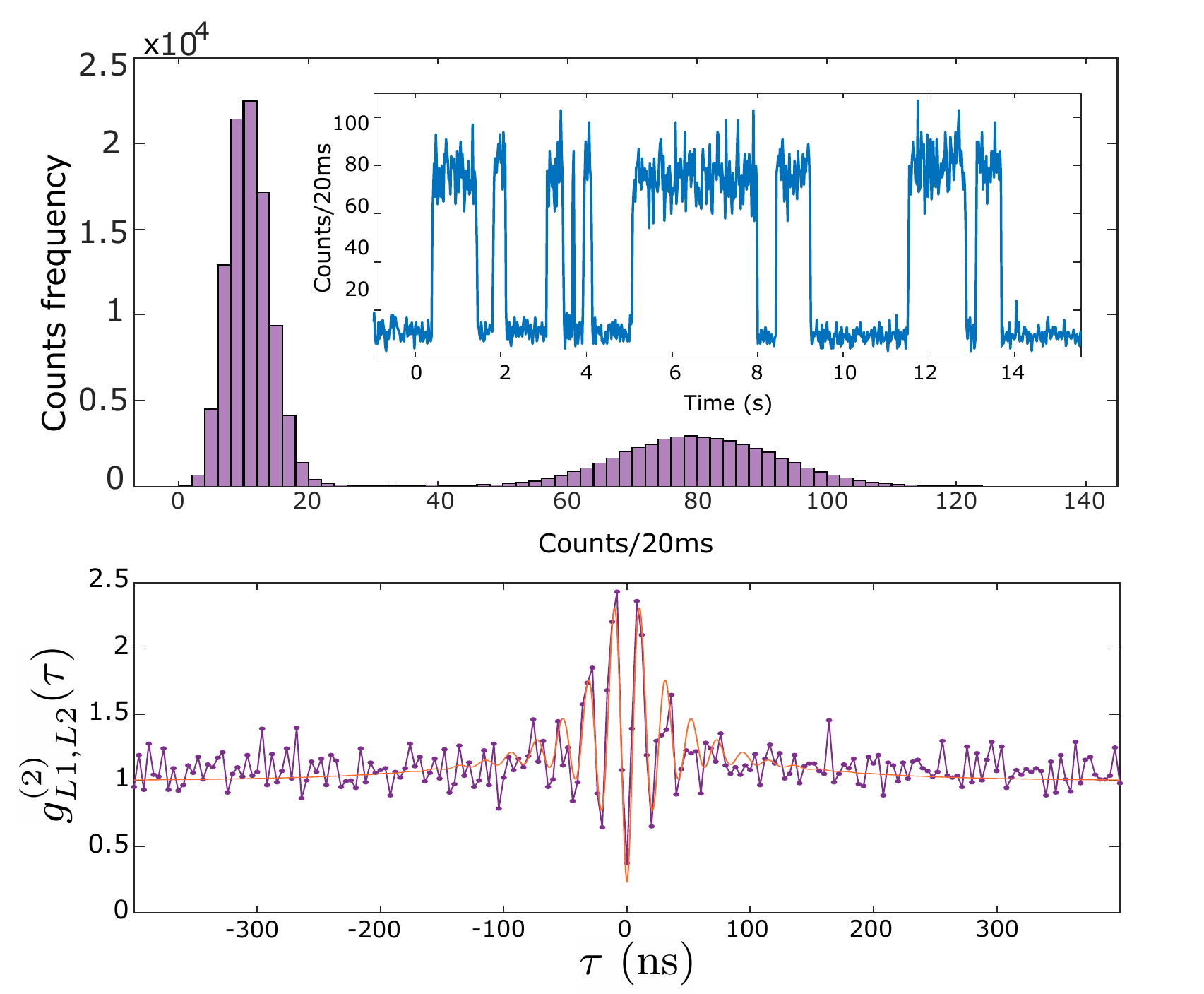} 
\caption{{Single-atom resonance fluorescence. Upper plot: time series (inset) and histogram (main graph) of collected fluorescence from L1 as single atoms enter and leave the trap with MOT and FORT in continuous operation, see text for details.   Lower plot: normalized cross-correlation $g^{(2)}_{L1,L2}(\tau)$ between collection channels L1 and L2.  Points show data, red curve shows a fit} {with $g^{(2)}_{L1,L2}(\tau)  = 1 + A \exp[-|\tau|/\tau_1]\cos(\Omega' t) + B\exp[-|\tau|/\tau_2]$ with fitting parameters $\Omega'=2 \pi \times \SI{7.57}{\mega\hertz}$ $\tau_1=\SI{28.6}{\nano \second}$ $\tau_2=\SI{90.9}{\nano \second}$ $A=1.26$ and $B=0.48$.}}
\label{fig:atom20s}
\end{center}
\end{figure}


\subsection{Fluorescence collection}
\label{sec:Collection}

The fluorescence collected by each lens is sent to a different channel of an avalanche photodiode detector (APD). APD counts in each channel are counted by an Arduino Due microcontroller and typically binned into \SI{20}{\milli\second} time bins.  A representative signal is shown in the inset of the upper plot of \autoref{fig:atom20s}. This shows a random telegraph signal, i.e., stochastic switching between just two signal levels, corresponding to the zero-atom and one-atom conditions. The main figure of the upper plot shows a histogram of the counts of this telegraph signal for a measurement of \SI{2700}{\second} duration.  It is clear that counts corresponding to zero atoms are well distinguishable from the counts corresponding to one atom in the trap. Due to LACs, larger atom numbers are not observed. We use this real-time telegraph signal for fine alignment of the collection fibers to the atom. The clear gap in counts allows us to perform sequence measurements triggered by the presence of an atom in the trap. The lower plot in \autoref{fig:atom20s} shows the normalized cross-correlation $g^{(2)}_{L1,L2}(\tau)$ of the signals collected via L1 and L2. Antibunching, i.e. $g^{(2)}_{L1,L2}(0)<1$,  indicates the presence of not more that one atom at a time in the trap. 

\begin{figure*}[t]
\begin{center}
\includegraphics[width=0.98 \textwidth]{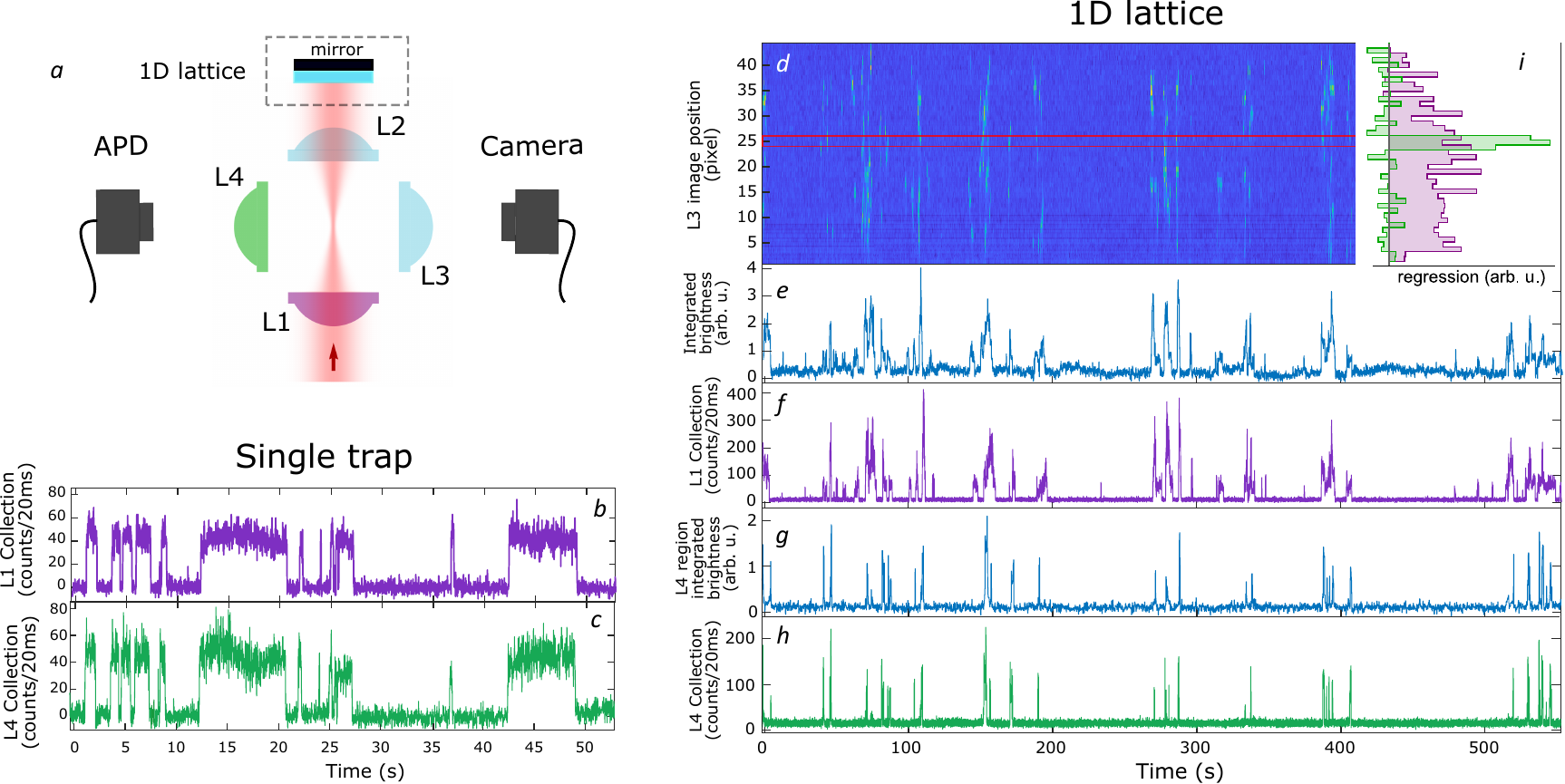}
\caption{{ Site-specific collection of light from a  simple FORT and a 1D lattice. \textbf{a}: Geometry of the trap, collection, and imaging optics.  \textbf{b} (\textbf{c}):  fluorescence signals collected by L1 (L4) of the simple FORT, i.e., with no retro-reflected beam, as seen on APDs.  \textbf{c}: spatially-resolved fluorescence over time from a continuously-loaded 1D optical lattice, imaged through L3 on an camera.  Vertical axis shows axial position in the lattice in pixels, at a magnification of \SI{1}{pixel\per\micro\meter}, horizontal axis shows time of acquisition. Colors indicate fluorescence intensity (arb. u.) integrated over a stripe of transverse dimension about the trap axis, increasing from dark to light. The same fluorescence signal is shown integrated over the length of the lattice in \textbf{e} and over the two pixels between red lines in \textbf{f}.  \textbf{g} (\textbf{h}): Single-mode fluorescence collection by L1 (L4).  \textbf{i} Contribution of different lattice locations (vertical axis, pixels on same scale as \textbf{d}) to the L1 (purple) and L4 (green) signals (horizontal axis). Values determined by linear regression, i.e. least-squares fit of a linear combination of camera pixel signals to the L1 and L4 APD signals.  } } 
\label{fig:RightAngleLattice}
\end{center}
\end{figure*}

\section{Trapping and collection in a right-angle geometry}
\label{sec:RightAngle}
{The selectivity in the collection at a right-angle to the trap axis is one of the advantages for the MCG, and provides more access channels when working in the single atom regime. To illustrate this, we produced a 1D optical lattice potential by reflecting the FORT light back through lens L2 in order to create a standing wave, as shown in \autoref{fig:RightAngleLattice}\textbf{a}. The input FORT power is reduced to \SI{2.5}{\milli\watt} to partially compensate the intensity boost implied by the standing wave geometry.  Atoms were randomly loaded from the free-running MOT into the lattice, and their fluorescence recorded with a camera via lens L3.  Simultaneously, light collected by L1 (along the lattice axis) and L4 (at a right angle) were coupled into single-mode fibers and detected with APDs.
{From the recorded video, the brightness of the images was integrated over a rectangular region $3$ pixels high (out of the plane of the four lenses) and $2$ or $45$ pixels long (along the $z$ axis) for L4 or L1, respectively, to obtain  \autoref{fig:RightAngleLattice}\textbf{e} and \textbf{g}, respectively. In \autoref{fig:RightAngleLattice}\textbf{e}-\textbf{h}, it is possible to compare APDs collection with the integrated intensity of the camera images as a function of time. In \autoref{fig:RightAngleLattice}\textbf{i} shows the contribution to the fluorescence collected by L1 and L4 as a function of position in the lattice. The right-angle collection is strongly correlated with a region of length \SI{2}{\micro\meter} in the longitudinal direction, covering $\approx 4$ lattice sites. The axial collection, in contrast, is correlated with the integrated intensity of the image as a whole.}   For comparison, \autoref{fig:RightAngleLattice}\textbf{b} and \textbf{c} show collection with lenses L1 and L4 with the single trap described in \autoref{sec:Overview}. In this condition, collection in the two directions is strongly correlated because each trapped atom explores the entire trap volume. Each channel presents a good signal-to-noise ratio.}


\section{Trap characterization} 
\label{sec:Characterization}
In this section, we report characterization of the main trap parameters.

\subsection{Occupancy and loading rate}
With the MOT running, loss of an atom from the FORT is most likely by LAC with the next atom to fall into the FORT.  For this reason, the trap occupancy is approximately 50\%, with the loading rate being nearly equal to the loss rate, as shown in \autoref{fig:atom20s} and \autoref{fig:RightAngleLattice}.  The loading rate can be controlled via the overlap of the MOT with the FORT, using the MOT compensation coils to displace the MOT.

\begin{figure}[t]
\begin{center}
\includegraphics[width=0.8\columnwidth]{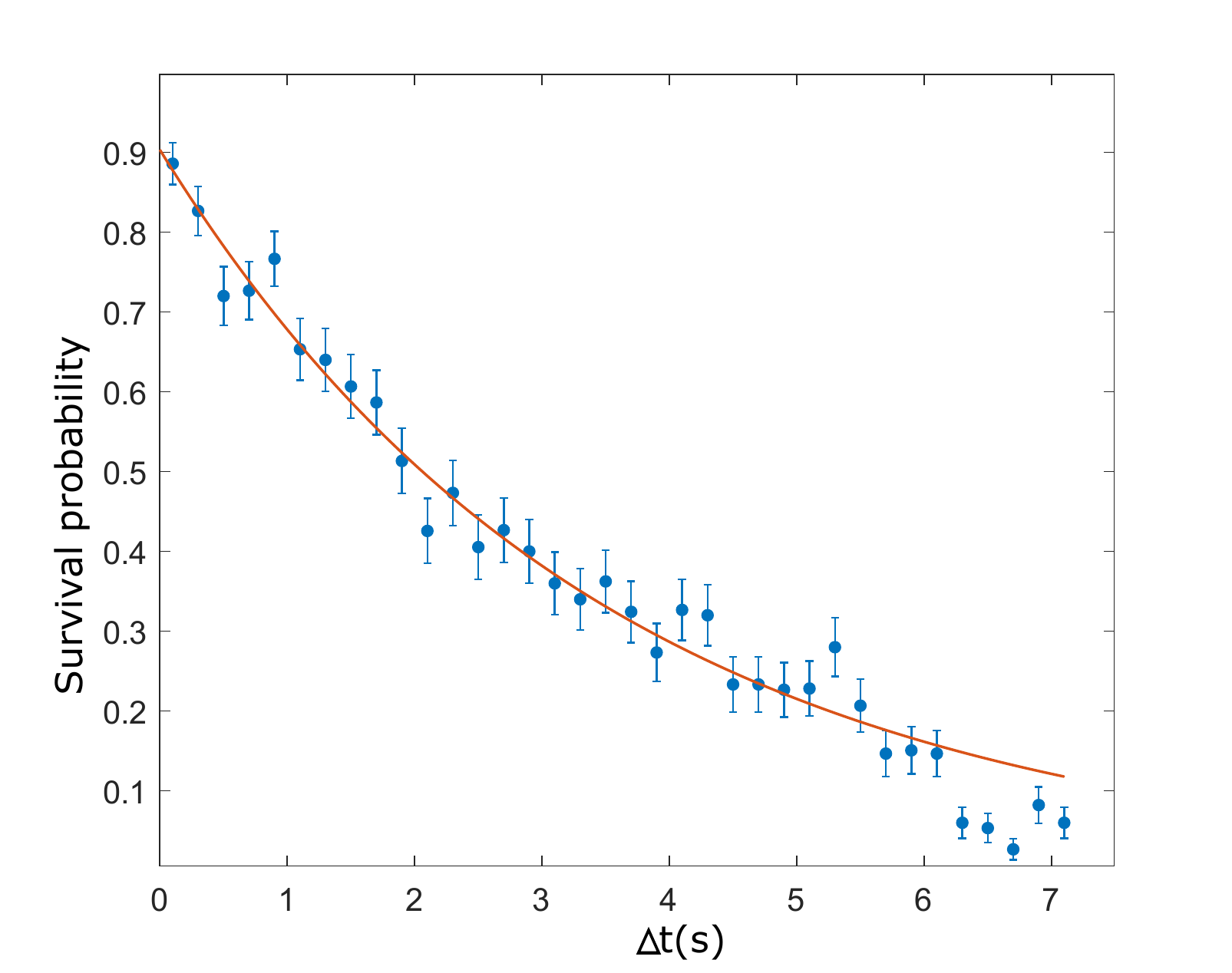}  
\caption{Persistence of a trapped atom in the FORT as a function of hold time $\Delta t$.  After detection of an atom by fluorescence, the MOT beams are turned off and the magnetic field gradient reduced, to prevent capture of a second atom in the FORT.  After at time $\Delta t$, the MOT beams are restored, and the presence or absence of the atom inferred from the fluorescence it produces.  Each point shows the average of 150 trials, error bars show $\pm$ one standard error, assuming binomial statistics. Line shows exponential fit with $1/e$ lifetime \SI{3.5+-0.3}{\second}.}
\label{fig:LifeTime}
\end{center}
\end{figure}

\subsection{Trap lifetime}
\label{sec:LifeTime}

By turning off the MOT beams when an atom's fluorescence is detected on the APD, it is possible to trap and hold an atom in the FORT without loss by LAC.  In this situation atoms can still be lost by collisions with background gas in the vacuum chamber, and by heating from stray light, scattering of the FORT beam, or FORT power or pointing fluctuations. The lifetime of an atom due to these effects was measured, with results shown in \autoref{fig:LifeTime}. The observed lifetime of \SI{3.5+-0.3}{\second} is typical in our setup. The lifetime decreases with increasing pressure in the vacuum chamber, for example when dispensers are heated to release Rb. This suggests that the loss is principally from collisions with background Rb atoms.

\begin{figure}[t]
\begin{center}
\vspace{-9mm}
\includegraphics[width=1\columnwidth]{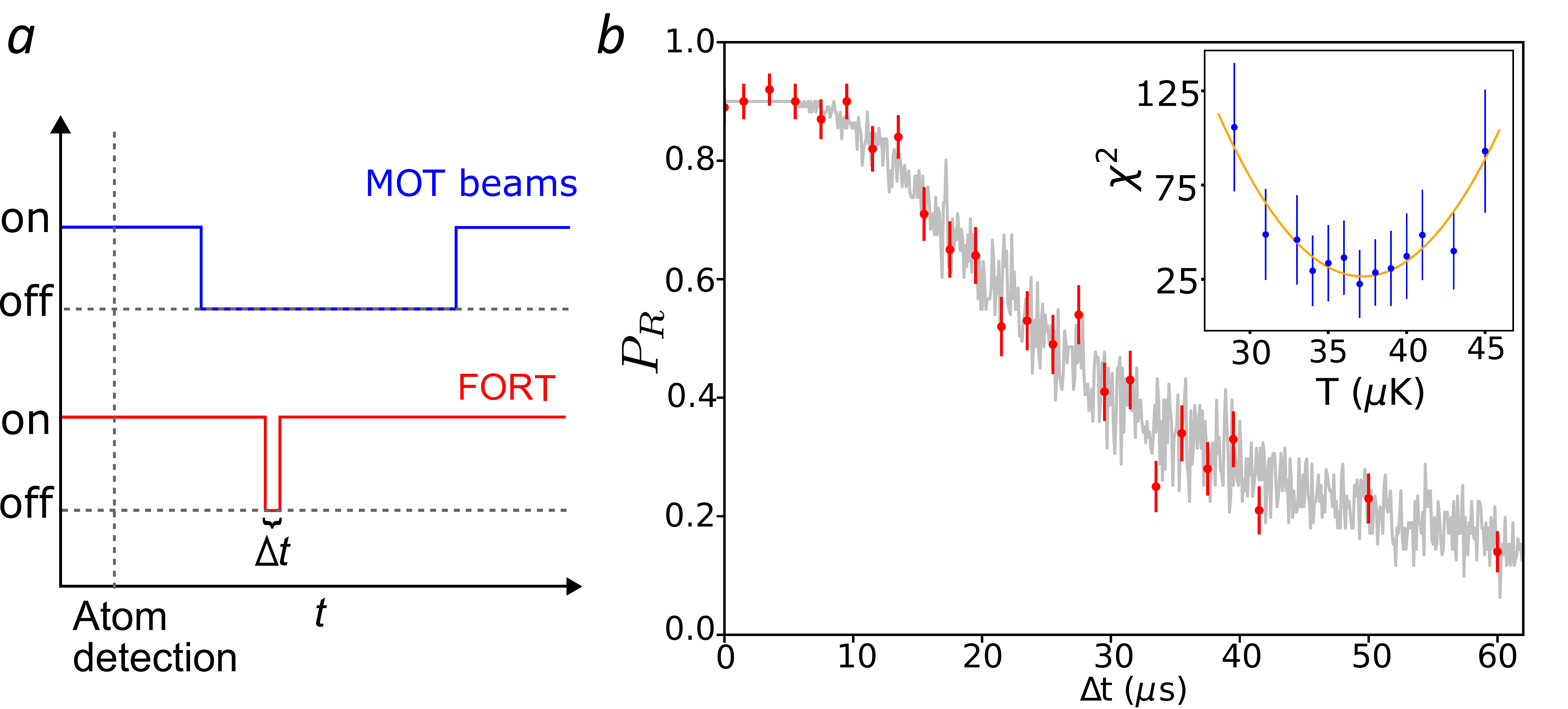}
\caption{Release and recapture measurement of atom temperature. \textbf{a}. Cooling, repumper and FORT beams temporal sequence (not to scale). \textbf{b}. Observed recaptured fraction $P_{R}$  as a function of the release time $\Delta t$ (red circles). Each point is the result of 100 trials. Error bars show $\pm $ one standard error of $P_R$ assuming a binomial distribution. Grey points show the recapture frequencies observe in a MC simulation with $T_{\rm atom}=\SI{37}{\mu\kelvin}$, and including a $\Delta t$-independent 11 percent probability of losing the atom between recapture and fluorescence detection. Inset: $\chi^2$ distance between data and MC simulation (blue circles) for different temperatures $T$. Error bars show $\pm$ one standard error of $\chi^2$ by propagation of error. A least-squares quadratic fit $\chi^2_{\rm fit}(T)$ (orange curve, see text) finds $T_{\rm atom}=\SI{37\pm 2}{\mu\kelvin}$.}
\label{fig:Temperature}
\end{center}
\end{figure}

\subsection{Atom temperature} 
\label{sec:Temperature}
We use the release and recapture method to determine the atom's temperature in the FORT, as illustrated in \autoref{fig:Temperature}\textbf{a}. We follow the protocol and analysis described in \cite{TuchendlerPRA2008}. The MOT and FORT are run until an atom is detected by its resonance fluorescence, as described above.
Repumper and cooler beams are then turned off and the MOT magnetic gradient reduced to prevent a second atom from falling into the trap. The FORT is then turned off for a time ${\Delta}t$, during which the atom can escape the FORT by ballistic motion. We then turn on the FORT, wait \SI{100}{\milli\second} and turn on the MOT beams. A recaptured atom is detected by the fluorescence it produces in this last phase. We repeat this sequence 100 times for each value of ${\Delta}t$. In \autoref{fig:Temperature}\textbf{b} we show the recaptured fraction $P_R(\Delta t)$ for typical conditions. 

We compare the experimental observations against a Monte Carlo (MC) simulation of the atom's probability to be recaptured.  In this simulation we assume that, at the moment the FORT is turned off, the atom's position is gaussian-distributed about the trap center, with zero mean and variances $\langle \Delta x^2 \rangle = \langle \Delta y^2 \rangle =k_{B}T/(m_{87}\omega_{r}^2)$ and $\langle \Delta z^2\rangle=k_{B}T/(m_{87}\omega_{z}^2)$, which follow from the equipartition theorem under the potential in the harmonic approximation.  We assume the atom's momentum distribution has zero mean and variance $\langle {\Delta}v_{x,y,z}^2\rangle ={k_{B}T/m_{87}}$, which describes the Maxwell-Boltzmann distribution. We then compute the evolved position ${\bf x}_f = {\bf x}(t = \Delta t)$ and velocity ${\bf v}_f = {\bf v}(t = \Delta t)$ after ballistic flight under gravity for time $\Delta t$, and the resulting total energy $E_T \equiv m_{87} {\bf v}_f^2/2 + U_{\rm FORT}({\bf x}_f)$ when the FORT is turned on at time $\Delta t$.  If  $E_T<0$, the atom is considered recaptured. 

For given $T$ and $\Delta t$, we repeat this sequence 100 times to find the recaptured fraction $f_R(T,\Delta t)$. To compare the simulation and experimental results, we calculate $\chi^2(T) = \sum_{\Delta t}{[f_{\rm R}(T, \Delta t)-P_{R}(\Delta t)]^2/\sigma^2(\Delta t)}$, where $\sigma(\Delta t)$ is the standard error of $P_{R}(\Delta t)$. 
As shown in \autoref{fig:Temperature}\textbf{b} (inset), we compute $\chi^2(T)$ for several $T$ and fit, by least squares, a quadratic function which we denote $\chi^2_{\rm fit}(T)$. The minimum of  $\chi^2_{\rm fit}(T)$ is taken as the best-guess temperature $T_{\rm atom}=\SI{37\pm 2}{\mu\kelvin}$, with uncertainty $\sqrt{2 \left \lfloor \partial^2 \chi^2_{\rm fit}(T)/\partial^2T \right\rfloor^{-1}}$ , where  $\left\lfloor \partial^2 \chi^2_{\rm fit}(T)/\partial^2T \right\rfloor$ is the 1-$\sigma$ lower confidence bound on $\partial^2 \chi^2_{\rm fit}(T)/\partial^2T$ \cite{BevingtonRobinson2003}. We note that $T_{\rm{atom}}\ll U_0/k_B \approx \SI{780}{\micro\kelvin}$, which justifies the harmonic approximation to the trapping potential.

\begin{figure}[t]
\begin{center}
\vspace{3mm}
\includegraphics[trim= 0 0 0 0,width=1\columnwidth]{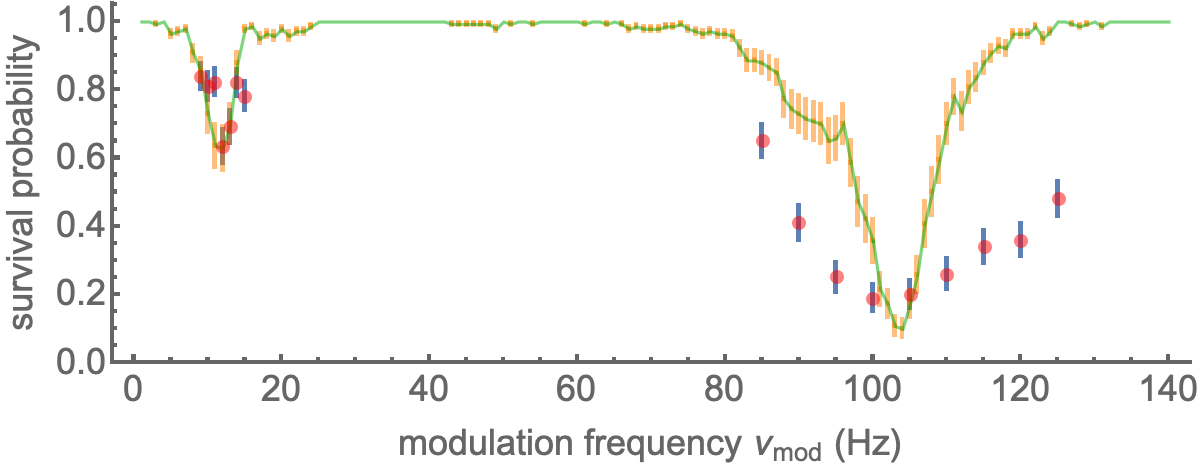} 
\caption{Parametric excitation of a single trapped atom, as described in the text.  Graph shows survival probability versus modulation frequency $\nu_{\rm mod}$.  Data (simulation) are shown as red (green) points. Error bars indicate $\pm$ one standard error assuming binomial statistics.}
\label{fig:TrapResonance}
\end{center}
\end{figure}

\subsection{Parametric resonances and trap frequency}
\label{sec:frequency}
Parametric excitation, in which the FORT power is modulated to excite parametric resonances in the atomic motion, is widely used to characterize the trap frequencies in optically-trapped atomic gases  \cite{WuJAP2006, ScheunemanRPA2000}.  With ensembles the heating rate, and thus the rate of loss from the trap,  shows resonances at specific frequencies.  In the harmonic approximation, these occur at double the trap frequencies, due to the even symmetry of the perturbation to the potential.  Corrections due to trap anharmonicity have been studied  \cite{WuJAP2006} and the technique has been applied to single atoms \cite{ShihPRA2013}.  

To measure these parametric resonances we used the following sequence: after loading an atom, we blocked the cooler light, leaving on the FORT and repumper beams, so the atom remained in the now-dark $F=2$ manifold. We then modulated the FORT power for time $t_{\rm mod}$ at a modulation frequency $\nu_{\rm mod}$ with a depth of modulation of $\approx \SI{20}{\percent}$. The power modulation was accomplished by sinusoidally modulating amplitude of the RF that drives the FORT acousto-optic modulator and thus the power of the first diffraction order into a single-mode fiber that leads to the experiment.  

Following the trap modulation, we checked for the presence of the atom by turning on again the cooler and collecting fluorescence.  We repeated this process for 100 atoms for values of $\nu_{\rm mod}$ near the second harmonic of the predicted \SI{6.7}{\kilo\hertz} longitudinal and  \SI{56}{\kilo\hertz} transverse trap frequencies in the harmonic regime.  In  the modulation was maintained for \SI{30}{\milli\second} in the lower-frequency range and \SI{150}{\milli\second} in the higher.  Results are shown in \autoref{fig:TrapResonance}, with  resonances at $\approx \SI{12}{\kilo\hertz}$ and $\approx \SI{100}{\kilo\hertz}$, about \SI{10}{\percent} lower than expected based on the trap frequencies previously estimated.  We note that anharmonicity has not been taken into account, and can be expected to shift the parametric resonances to lower frequency relative to the second harmonics of the trap frequencies \cite{WuJAP2006}. Also shown are the results of a MC simulation, in which atoms drawn from a Boltzmann distribution as in \autoref{sec:Temperature} are allowed to evolve under the modulated potential plus a Langevin term describing isotropic noise from scattering of background and FORT photons.  A reasonable agreement can be obtained with modulation depth of $\approx \SI{36}{\percent}$, heating rate $\approx \SI{400}{recoil\per\second}$, and trap waist $w_0 \approx \SI{1.5}{\micro\meter}$.

~

\section{Conclusion}
\label{sec:Conclusion}
{We have described a system for stable, long-term trapping and cooling of single $^{87}$Rb atoms at the center of a Maltese cross geometry optical system of four high-NA aspheric lenses in vacuum. The system gives high-NA access to the common focal region, allowing selective identification of trap regions in all three dimensions.  We have studied the principal characteristics of this trapping system, including the loading dynamics, trap lifetime, visibility of single-atom signals, in-trap atom temperature and parametric excitation spectrum.  We find trap performance comparable to what has been reported for single-atom traps with one- or two-lens optical systems. We conclude that the multi-directional high-NA access provided by the Maltese cross geometry can be achieved while preserving other trap characteristics such as lifetime, temperature, and trap size.
}
\\
\section*{Acknowledgments}
The authors thank Ludovic Brossard and Antoine Browaeys for helpful discussions. This project was supported by Spanish Ministry of Science projects OCARINA (Grant No. PGC2018-097056-B-I00), Q-CLOCKS (Grant No. PCI2018-092973), and ``Severo Ochoa'' Center of Excellence CEX2019-000910-S
Generalitat de Catalunya through the CERCA program; 
Ag\`{e}ncia de Gesti\'{o} d'Ajuts Universitaris i de Recerca Grant No. 2017-SGR-1354;  Secretaria d'Universitats i Recerca del Departament d'Empresa i Coneixement de la Generalitat de Catalunya, co-funded by the European Union Regional Development Fund within the ERDF Operational Program of Catalunya (project QuantumCat, ref. 001-P-001644); Fundaci\'{o} Privada Cellex; Fundaci\'{o} Mir-Puig;  17FUN03-USOQS, which has received funding from the EMPIR programme co-financed by the Participating States and from the European Union's Horizon 2020 research and innovation programme.
\section*{References}
\bibliographystyle{./apsrev41nourlmod}
\bibliography{../biblio/biblio}

\end{document}